\documentstyle[multicol,aps,psfig,epsf]{revtex}
\newcommand{\lan}{\langle}
\newcommand{\ran}{\rangle}
\begin{document}
\twocolumn[\hsize\textwidth\columnwidth\hsize\csname 
 @twocolumnfalse\endcsname
\title{Aging and multiscaling in out of equilibrium dynamical processes 
of granular media}
\author{Mario Nicodemi and Antonio Coniglio}
\pagestyle{myheadings}
\address{Dipartimento di Fisica, Universit\'a di Napoli ``Federico II'',
Unit\`a INFM and INFN Napoli \\
Mostra d'Oltremare, Pad. 19, 80125 Napoli, Italy}

\maketitle
\date{\today}
\begin{abstract}
In the framework of recently introduced frustrated lattice gas models, 
we study the out of equilibrium dynamical processes during the compaction 
process in granular media. We find irreversible-reversible cycles in 
agreement with recent experimental observations. Moreover
in analogy with the phenomenology of 
the glass transition we find aging effects during the compaction process
In particular we find that the two time density 
correlation function $C(t,t')$ asymptotically scales as a function of the 
single variable $\ln(t')/\ln(t)$. This result is interpreted  
in terms of multiscaling properties of the system.
\end{abstract}
%\smallskip
%{\small Key words: Granular Media, Glasses, Slow Relaxation}
%\smallskip
%\newpage
\vskip2pc]

%\section{Introduction}

The experimental study of dynamic processes in granular media \cite{JNBHM} 
has recently revealed the presence of interesting behaviours. 
Under tapping dry granular media reach very slowly a more
compact state which is well fitted by
a logarithmic relaxation \cite{Knight}. More recently Novak et al.\cite{Novak} 
have also shown that such materials exhibit 
non trivial irreversible-reversible cycles.  
These phenomena stem from slow relaxation processes due to large
``cooperative rearrangements'' of many particles.
In such a perspective granular materials share features of 
thermal systems such as glasses or spin glasses which are also 
characterized by extremely long relaxation times and 
the presence of 
irreversible-reversible cycles \cite{angell,BCKM}. 

In this paper in the framework of recently introduced 
microscopic models \cite{CH,NCH,Caglioti}, we reproduce 
the irreversible-reversible cycles of Novak et al.
and investigate the effect of the ``cooling''  rate 
on the compaction process. We find a behaviour which is 
strongly reminiscent of the phenomenology of the
glass  transition. Finally we study 
the non equilibrium time dependent density density autocorrelation function
and found aging effects typical of glassy systems.
These results suggest that similar effects could be also found in real 
experiments.

%\section{Frustrated lattice gas models}

In dynamical processes of granular media a crucial role is played 
by geometric frustration (originated by steric hindrance between interlocked 
neighboring grains) which induce the necessity of large scale cooperative 
rearrangements for relaxation. Based on these concepts two kinds of 
frustrated lattice gas model were introduced \cite{CH,NCH,Caglioti}. 
Both models showed logarithmic compaction, segregation and other 
phenomena typical of granular media under shaking. 

\begin{figure}[ht]
\centerline{\psfig{figure=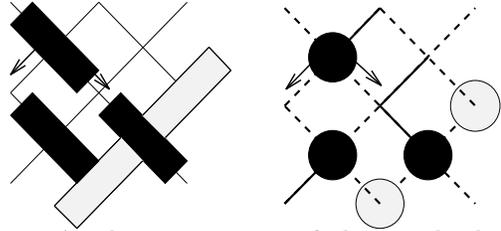,height=3.0cm,width=6.5cm,angle=-90}}
%\vspace{0.5cm}
\caption{A schematic picture of 
the two kind of frustrated lattice gas models described in the 
text. {\em Left:} the Tetris model. 
{\em Right:} the Ising frustrated lattice gas, IFLG.
Straight and dashed lines represent the two kind of interactions
$\epsilon_{ij}=\pm 1$. 
Filled circles are present particles with ``orientation" 
$S_i=\pm 1$ (black/white).
} 
\label{lattice}
\end{figure}

These models consist of a system of particles which occupy the sites 
of a square lattice tilted by $45^{\circ}$ (see Fig.\ref{lattice}). 
Particles are characterized by an internal degree of freedom, 
$S_i=\pm 1$, corresponding for instance to two typical orientations 
of grains on 
the lattice. Two nearest neighbor sites can be both occupied only if  
the particles have the right reciprocal orientation, so that they do 
not overlap, otherwise, due to excluded volume, they have to move away.
In absence of vibrations the particles are subject only to gravity and they 
move downwards always fulfilling the non overlap constraint. 
The effect of vibration is introduced by allowing the particles to diffuse 
with a probability $p_{up}$ upwards and a probability $p_{down}=1-p_{up}$
downwards. An important parameter governing the dynamics is the 
adimensional parameter $\Gamma\equiv -1/\ln(x_0)$, 
with $x_0=p_{up}/p_{down}$,
which is related (see below) to the effective temperature of the system
and consequently plays the same role as the amplitude of the vibration intensity
in the experiment of Novak et al. \cite{Novak}.

Such models can be described in terms of the following lattice gas 
Hamiltonian (see \cite{NCH,Caglioti}) in the limit $J\rightarrow\infty$: 
\begin{equation}
H=J\sum_{\langle ij\rangle } f_{ij}(S_i,S_j)n_i n_j 
\label{H}
\end{equation}
Here $n_i=0,1$ are occupancy variables,
$S_i=\pm 1$ are spin variables, associated to the two orientations of the
particles, $J$ represents the infinite repulsion
felt by the particles when they have the wrong orientations and 
$f_{ij}(S_i,S_j) = 0$ or $1$ depending whether the configuration 
$S_i,S_j$ is right (allowed) or wrong (not allowed). 

The choice of $f_{ij}(S_i,S_j)$  depends on the particular model. 
The Tetris \cite{Caglioti} model is made of elongated particles 
(see Fig.~\ref{lattice}), 
which may point in two 
directions coinciding with the two lattice bond orientations. 
In this case $f_{ij}(S_i,S_j)$ is given by 
$f_{ij}(S_i,S_j)=1/2(S_i S_j -\epsilon_{ij} (S_i+ S_j) +1)$, where 
$\epsilon_{ij}=+ 1 $ for bonds along one direction of the lattice and 
$\epsilon_{ij}=- 1 $ for bonds on the other. 
This hamiltonian model has an ordered ``antiferromagnetic'' ground 
state, and its dynamics has the crucial constraint that 
particles can flip their ``spin" only if 3 of their own 
neighbors are empty.

A real granular system may contain more disorder 
due to a wider shape distribution or to the absence of a lattice. 
Therefore in a more realistic model 
the number of internal states is $q>2$
($S_i=1,2...q$)
and the function $f_{ij}(S_i,S_j)$ is zero only for allowed 
nearest neighbor configurations. 

However to simplify the model the number of states was still kept $q=2$
and the randomness was taken into account by introducing
random quenched variables corresponding to the freezing of some degree
of freedom in the high density regime. Thus 
an Ising  frustrated lattice gas model (IFLG)
was proposed \cite{NCH} in which $f_{ij}(S_i,S_j)$ was given by
$f_{ij}(S_i,S_j)=1/2(\epsilon_{ij} S_i S_j -1)$
and $\epsilon_{ij}=\pm 1$ are 
quenched random interactions associated to the
bonds of the lattice. 

The Hamiltonian~(\ref{H}) is without gravity. In presence of gravity there is 
an extra term $g\sum_i n_iy_i$ 
where g is the gravity and $y_i$ is the ordinate of the particle $i$. 
The temperature T 
is related to the ratio $x_0=p_{up}/p_{down}$  via $e^{-2g/T} = x_0$, 
(notice $\Gamma=T/2g$ ).

Interesting enough the two extreme models, the Tetris and the IFLG, 
show similar behaviour. This suggests that the
results found are rather robust and will not depend much on
the details of the model. In particular under tapping they
reproduce the
logarithmic behaviour in agreement with the experimental results
of  Knight et al.\cite{Knight}. 
Experimentally a ``tap" is the shaking of the container of the grains 
by vibrations of given duration and amplitude.
In our Monte Carlo simulations 
each single tap is  realized by letting the particle diffuse 
under the gravity by keeping $\Gamma=const.$ 
during the time interval of a tap, $\tau_0$, 
and then switching off the vibration by setting $\Gamma=0$ 
until the system reaches a static configuration. 
Time $t$ is measured in such a way that one unit corresponds to one single
average update of all particles and all spins of the lattice. 

In a previous paper \cite{NCH} we have performed on the IFLG a particular 
cycle sequence of taps to show the presence in granular media of 
hysteresis effect.
Recently also real experiments were performed 
on density relaxation in 
granular media under cyclic tapping \cite{Novak}, which show indeed 
the presence  of such hysteresis effect.
In order to compare better with the experimental data we have
simulated the same cycle of Novak et al. on the IFLG model. 
We have considered a system of size 
$30 \times 60$ (our data are robust to size changes), with periodic
boundary conditions along the x-axis and rigid walls at bottom and 
top. 
Our data are averaged over $8$ different 
lattice realizations each averaged over $30$ different noise realizations.

A starting particle configuration is
prepared by randomly inserting particles 
into the box from its top and then letting them fall down, with the
dynamics described above, until the box is filled. 
We performed cycles of taps in which the
vibration amplitude $\Gamma$ is varied at fixed
amplitude increment $\gamma=\Delta \Gamma/\tau_0$ 
holding constant their duration $\tau_0$. 
More precisely we performed 
a sequence of ${\cal{N}}$ taps, of amplitude 
$\Gamma_1 ... \Gamma_n ... \Gamma_{{\cal{N}}}$, 
from an initial amplitude $\Gamma_1=0$ to a
maximal amplitude $\Gamma_{max}=15$, then back to $\Gamma=0$ and 
then up again to $\Gamma_{{\cal{N}}}=\Gamma_{max}$.
After each tap we measure the static bulk density of the system 
$\rho(\Gamma_n)$ ($n$ is the $n$-th tap number).

\begin{figure}[ht]
\centerline{\psfig{figure=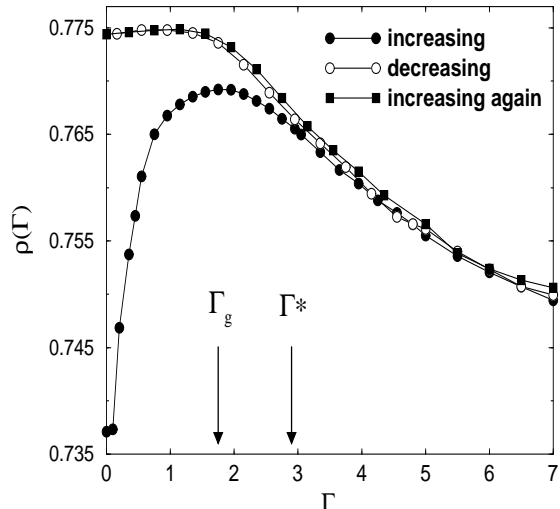,height=8cm,width=8cm,angle=-90}}
%\centerline{\psfig{figure=hyst_cycle_6_1.ps,height=5cm,width=5cm,angle=-90}}
%\vspace{0.5cm}
\caption{The static bulk density, $\rho(\Gamma)$, of the IFLG model 
(the Tetris gives analogous results) as a function of 
the vibration amplitude, $\Gamma$, in cyclic vibration sequences. 
The system is shaken with an amplitude $\Gamma$ which at first 
is increased (filled circles), then is decreased (empty circles) 
and, finally, increased again (filled squares) with a
given  ``annealing-cooling" velocity 
$\gamma\equiv \Delta\Gamma/\tau_0$
(at each value of $\Gamma$ the system has undergone a ``tap" of
duration $\tau_0=10^3$). Here we fixed $\gamma=1.25~10^{-3}$. 
The upper part of the cycle is approximately ``reversible" 
(i.e., empty circles and filled squares fall roughly on the 
same curve). 
The data compare rather well with the experimental data of 
Novak et al.. 
$\Gamma^*$ is approximately the point where the ``irreversible" and 
the ``reversible" branches meet. $\Gamma_g$ signals the 
location of a ``glass transition" (see Fig.\ref{den_hys_up_branch}).
} 
\label{den_hys}
\end{figure}

\begin{figure}[ht]
\centerline{\psfig{figure=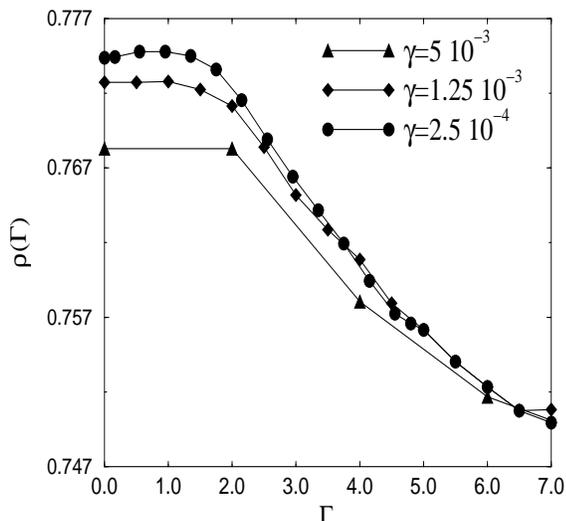,height=8cm,width=8cm,angle=-90}}
%\centerline{\psfig{figure=hyst_cycle_6_1.ps,height=5cm,width=5cm,angle=-90}}
%\vspace{0.5cm}
\caption{
As in Fig.\ref{den_hys}, we report 
the density, $\rho(\Gamma)$, as a function of 
the vibration amplitude, $\Gamma$ 
for three different values 
of the ``cooling'' velocity $\gamma$.
%$\gamma=5~10^{-3}$, triangles, $\gamma=1.25~10^{-3}$, diamonds, 
%$\gamma=2.5~10^{-4}$, circles). . 
For sake of clarity, we plot here only the 
descending reversible parts of the cycle. 
As in 
experiments on glasses a too fast cooling drives the system out of 
equilibrium. The position of the shoulder, $\Gamma_g(\gamma)$, 
in these curves schematically individuates a ``glass transition". 
} 
\label{den_hys_up_branch}
\end{figure}

Our results are qualitatively very similar to those reported in 
real experiments on dry granular packs \cite{Novak}. 
We find that when the system is successively shaken at increasing vibration amplitudes, 
the bulk density of the system typically grows and then decreases 
as shown in Fig.~\ref{den_hys}. However, 
when the amplitude of shaking decreases back, 
the density follows the same path only up to some value of $\Gamma$
and then deviates from it, in fact it does not bend down and keeps growing.
As in the experimental data 
the second part of the shaking cycle is approximately reversible 
(see Fig.~\ref{den_hys}). 
For such a reason these processes are called 
``irreversible-reversible" cycles \cite{Novak}.
Interestingly the reversible cycle is a monotonic function of the 
shaking amplitude (see Fig.\ref{den_hys_up_branch}). 

To study the dependence on the ``cooling" rate $\gamma$,
we repeated the tapping sequence for different values of the tap amplitude 
increment $\Delta \Gamma$ with a fixed tap duration $\tau_0$ 
(see Fig.\ref{den_hys_up_branch}). 
We find that the reversible branches have a common part 
for high values of $\Gamma$ while for small $\Gamma$ they 
splits in different curves 
depending on the cooling rate $\gamma$. The slower the cooling rate the 
higher is the 
the final density observed at the end of the descending part of the cycle.
One can schematically define the point, $\Gamma_g(\gamma)$, where 
the system freezes and goes out of equilibrium as the location of the 
``shoulder" in these ``reversible" branches (see Fig.\ref{den_hys} 
and \ref{den_hys_up_branch}). Thus, $\Gamma_g(\gamma)$, which 
depends on $\gamma$, corresponds to a ``glass transition". 
Notice that $\Gamma_g(\gamma)$ is usually different from 
the point, $\Gamma^*(\gamma)$, where the ``irreversible" and 
the ``reversible" branches meet (see Fig.\ref{den_hys}). 
However, as $\gamma$ gets smaller $\Gamma_g$ and $\Gamma^*$ become 
closer and they may coincide in the limit $\gamma\rightarrow 0$.

As in glasses the system gets out of
equilibrium due to the fact that the 
characteristic times of relaxation are much larger than the time $\tau_0$ 
involved in the experiment, but, for the same reason, the location of the 
path depends on the rate $\gamma$. 

The limit of $\gamma$ going to zero defines an ideal glass transition
amplitude  $\Gamma_0$. 
Using the analogy with the glass transition we expect 
a slow logarithm dependence of $\Gamma_g(\gamma)$ on $\gamma$ as in the 
glass transition.
Full details will be presented elsewhere 
\cite{NC_fut}. 
Experimental results in this direction will be also very interesting.

%\subsection{Correlation functions}

By further exploiting the analogy with the glass transition, we expect 
aging phenomena.
For instance if one keep shaking the system at a low 
fixed amplitude $\Gamma$ on the reversible branch, the system will 
slowly (logarithmically) approach an asymptotic value of the density, 
eventually on the equilibrium curve.
In order to further quantitatively characterize the out of equilibrium 
dynamics and aging phenomena in granular matter and make
quantitative predictions, in analogy with glassy system
we introduce a two time density-density correlation function ($t\ge t'$):
\begin{equation}
C(t,t')=\frac{\lan\rho(t)\rho(t')\ran-\lan\rho(t)\ran\lan\rho(t')\ran}
{\lan\rho(t')^2\ran-\lan\rho(t')\ran^2}
\label{cor}
\end{equation}
where $\rho(t)$ is the bulk density of the system at time $t$. 
In out of equilibrium $C(t,t')$ is a function of both times $t$ and $t'$
(at equilibrium just of $t-t'$). The aging properties of the system
are characterized by the specific scaling properties  
of $C(t,t')$ 

In order to study the system in a well defined configuration of its 
parameters, we evaluate $C(t,t')$ during a ``single tap": we prepare 
the system at $t=0$ by randomly pouring grains in the box from above 
as described before, 
then we start to shake it continuously and indefinitely with a 
given (small) amplitude $\Gamma$. We expect very similar results 
by considering, instead of a long tap, a series of short taps which 
is experimentally more convenient (as in ref.~\cite{Knight}). 
The data about $C(t,t')$ we present here are averaged, at least, 
over $8$ different lattice and $512$ different noise realizations. 

\begin{figure}[ht]
\centerline{\psfig{figure=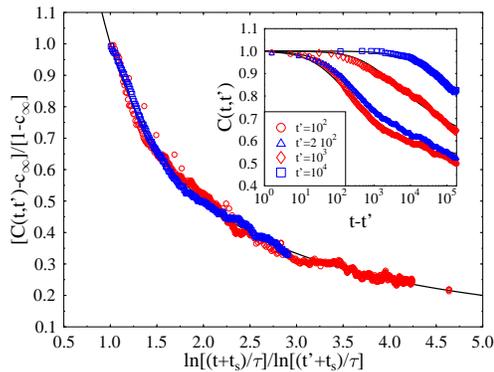,height=8cm,width=6cm,angle=-90}}
%\vspace{0.5cm}
\caption{ The two time density-density correlation function, 
$(C(t,t')-c_{\infty})/(1-c_{\infty})$, 
as a function of the scaling variable 
$\alpha=\ln[(t+t_s)/\tau]/\ln[(t'+t_s)/\tau]$. 
Scaled on the same master function are data from both models considered 
in the present paper (Tetris, squares, and IFLG, circles) 
and for vibration amplitudes $\Gamma=-1/\ln(x_0)$ with 
$x_0\in[10^{-4},10^{-1}]$. The master function is $1/\alpha$. 
{\em Inset:} The correlation $C(t,t')$ for the Tetris at $\Gamma=0.22$ 
(or $x_0=0.01$) as a function of $t-t'$ for four values of 
$t'=10^2, 2~10^2, 10^3, 10^4$. 
} 
\label{fig_cor_sca}
\end{figure}

At low $\Gamma$, a good fit for the two time 
correlation function, $C(t,t')$, on the whole five decades in 
time explored, is given by the following: 
\begin{equation}
C(t,t')=(1-c_{\infty}) \frac{\ln[(t'+t_s)/\tau]}{\ln[(t+t_s)/\tau]} 
+ c_{\infty}
\label{cor_sca}
\end{equation}
where $\tau$, $t_s$ and $c_{\infty}$ are fit parameters. 
Very interesting is the fact that the above behavior is found in both of 
our models (Tetris and IFLG).
The data for the two models, for several values of $\Gamma$, 
rescaled on a single universal master 
function, are plotted in Fig.~\ref{fig_cor_sca}. 
In particular, in the explored range of $\Gamma\in[0.11,0.43]$
(i.e., $x_0\in[10^{-4},10^{-1}]$), 
we found $\tau\sim e^{z/\Gamma}$ (with $z\sim 2$ in both the 
IFLG and the Tetris), the mark of activated dynamics 
($t_s(\Gamma)$ behaves approximately as 
$\tau$ as a function of $\Gamma$). 
The asymptotic value $c_{\infty}$ is difficult 
to determine with some precision: in the above $\Gamma$ range we 
evaluated approximately $c_{\infty}=0.2 \div 0.3$ for the IFLG model 
and $c_{\infty}=0.0 \div 0.2$ for the Tetris model.

Eq.~(\ref{cor_sca}) essentially states that for times long enough  
the correlation $C(t,t')$ is a function (linear) of the ratio
$\ln(t')/\ln(t)$. Such a scaling behaviour is known in 
others disordered systems like Random Ferromagnets or Random Fields 
models \cite{Bray} and has been proposed by Fisher and Huse droplets 
theory of finite dimensional spin glasses \cite{FH}. 
However, it seems to be different from other scaling functions 
proposed to fit experimental data in spin glasses \cite{BCKM}. 
All this shows the necessity of experimental confirmation of 
our results in the framework of granular media.

In any case, eq.~(\ref{cor_sca}) is in good agreement 
with a general multiscaling 
approach presented in Refs.~\cite{CM,CN_fut}. 
Whenever the characteristic relaxation times of the system becomes 
exceedingly large respect to those typical of the measurements, 
it is reasonable to assume scaling properties for the system. 
The most general scaling transformation which satisfies group properties
gives rise to a multiscaling form of the correlation function given by
\cite{CN_fut}:
\begin{equation}
C(t,t')={\cal{C}}(\alpha) t^{f(\alpha)}
\label{multisca}
\end{equation}
with a twofold possibility for $\alpha$: $\alpha=t'/t^z$ or 
$\alpha=\ln(t')/\ln(t)+O(1/\ln(t))$; here 
$\cal{C}(\alpha)$ and $f(\alpha)$ are two generic functions.
The assumption of multiscaling might thus naturally allow 
to explain the presence of large classes of 
universality found in scaling behaviour of apparently different systems.
The above case of eq.~(\ref{cor_sca}), observed in both Tetris and IFLG, 
corresponds to $\alpha=\ln(t')/\ln(t)$ and $f(\alpha)\sim 0$.

%\section{Conclusions}

In conclusion, in the framework of simple frustrated 
lattice gas models, we have studied 
the off equilibrium dynamics of slightly shaken 
granular materials. These models have previously shown to share 
many phenomena characteristic of granular media 
as logarithmic compaction or segregation. 
Here, we have studied irreversible-reversible cycles and found
good agreement with the experimental data on granular packs \cite{Novak}.
We have stressed the strong analogies with 
non equilibrium dynamic properties observed in 
glassy systems. 
In particular we 
evaluated the two time density correlation function $C(t,t')$,
which are found asymptotically to be function of the single ratio 
$\alpha=\ln(t')/\ln(t)$. This results observed in different models, and 
amenable to experimental 
check, may be interpreted by assuming multiscaling properties of the system.

We thank INFM-CINECA for computer time on Cray-T3D/E.

\end{document}